\documentstyle[preprint,epsf,aps]{revtex}
\begin{document}
\title{Reply to Cen LiXiang, Li XinQi, and Yan YiJing\\ On the nonadiabatic
conditional geometrical phase with NMR 
}
\author{Wang Xiang-bin\thanks{email: wang$@$qci.jst.go.jp} }

\maketitle 
Much long before the appearing time of the Comment\cite{cen}, 
the main issue addresed there had been resolved already.
Our correction to Eq.(11) in the Letter\cite{wang}
first appeared in quant-ph/0101038(v4 and onwards), 
then appeared again in the extended version, 
quant-ph/0108111, and were finalized  in 
ref\cite{ewang}(Erratum: PRL, 88, 179901(2002)).
Surprisingly, in the Comment, Cen {\it et al} have 
neither mentioned the fact that actually they are not the first
discoverer of  the issue nor mentioned the fact that Eq.(11) 
in the Letter\cite{wang}
has been replced by a new one in the Erratum\cite{ewang}, though they had 
been surely aware of these facts long time ago. 
It's W. A. Munro, T. P. Spiller, and A. Nazir who
first spotted the issue more than one year ago. Munro {\it et al} are acknowledged in our Erratum\cite{ewang}.

 Although we are not in a position to answer  Cen, Li and Yan's 
question to other persons' work\cite{zhu} in their Comment, we believe that
the dynamical phase in Zhu and Wang's work\cite{zhu} can be removed 
either in a way similar to 
that in ref\cite{ewang} or in a different way. Taking this point into consideration, we believe the main idea in Zhu and Wang's paper works. 

In summary, as Cen, Li, and Yan had  clearly known it 
long before, they are definitely not the first 
discoverer on the problem of Eq.(11) in our Letter. 
The problem of Eq(11) in our Letter had been pointed out and
corrected already long before their re-discovery. The information offered by
Cen, Li, and Yan in the Comment\cite{cen} are selective and misleading.    
    
\end{document}